\documentclass{appolb}
\usepackage{graphicx}
\usepackage{amsmath}

\begin{document}
\title{Next-to-Leading Order virtual correction to Higgs-induced DIS%
\thanks{Presented at ``Diffraction and Low-$x$ 2022'', Corigliano Calabro (Italy), September 24-30, 2022.}%
}
\author{Anna Rinaudo
\address{Dipartimento di Fisica, Università di Genova}
}
\maketitle
\begin{abstract}
We calculate the Next-to-Leading Order (NLO) virtual correction to the Higgs-induced DIS coefficient function in the infinite top-mass limit. Since we want to use this result in the framework of kt-factorization to resum small-x logarithms up to Next-to-Leading-Logarithm (NLL), we work in light-cone gauge and we keep the incoming gluon off-shell. This choice raises many challenging points like the presence of spurious singularities and a different definition for the UV-counterterms. This calculation is a necessary ingredient for the coefficient function that will be used to resum up to NLL small-x logarithms for this process.
\end{abstract}

\section{Introduction}

Using the $k_t$-factorization theorem, we can write the resummed coefficient function for a process as \cite{Bonvini:2016wki}
\begin{equation}
    C(N,\alpha_s) = \int d k_t^2\,\mathcal{C}(N,k_t^2,Q^2,\alpha_s)\,\mathcal{U}(N,k_t^2,Q^2).
\end{equation}
The first term in the convolution is the off-shell coefficient function for the process that in axial gauge is free form small-x logarithms. The other term is a universal factor that, computed at the required logarithmic accuracy, resums the small-x logs. This formula works well at Leading-Logarithmic (LL) accuracy \cite{Bonvini:2016wki} and we want to test it at Next-to-Leading-Log (NLL) accuracy. To do so, we have to compute the off-shell coefficient function at Next-to-Leading Order (NLO) and the factor $\mathcal{U}$ at NNL. 

In this work, we present a preliminary result for the NLO virtual contribution to the coefficient function for a specific process, the Deep-Inelastic-Scattering (DIS) initiated by a Higgs boson. We chose this process because it is simple enough to allow us to understand some technical issues like the requirement to work in axial gauge. 

\section{Higgs-induced Deep-Inelastic-Scattering}

\begin{figure}[h]
    \centering
    \includegraphics[width = 0.25 \linewidth]{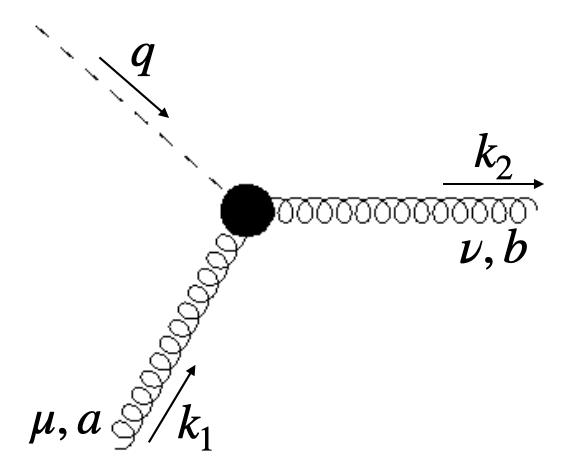}
    \caption{Higgs induced DIS.}
    \label{fig:HiggsDIS}
\end{figure}

The process we are interested in is
\begin{equation}
  g^{(*)} (k_1) + H (q) \rightarrow g (k_2), 
  \label{eq:Higgs-DIS}
\end{equation}
where the initial state gluon is off-shell. The Feynman rule for the lowest order coupling between Higgs and gluons is
\begin{equation}
	M^{\mu \nu}_{ab}(k_1, k_2) =  \mathrm{i} \, c \left( k_1^\nu k_2^\mu  -  g^{\mu \nu}k_1\cdot k_2 \right) \delta_{ab}, 
\end{equation}
where $c = \frac{\alpha_s \sqrt{G_F \sqrt{2}} }{3 \pi}$ and $a$, $b$ are the colours of the gluons.
The momenta of the involved partons are shown in Fig.\ref{fig:HiggsDIS}.
Following the $k_t$-factorization procedure, we identify the offshellness of the incoming gluon with its transverse component. In particular, we define 
\begin{equation}
    k_1^\mu = k^\mu + k_t^\mu,
\end{equation}
so that $k\cdot k_t = 0$ and $k_1^2 = -\textbf{k}_t^2$. 

We compute the one-loop contribution to this process in light-cone gauge. This means that the gauge vector $n^\mu$ is defined so that $A\cdot n = 0$ and $n^2=0$.

To compute the coefficient function for the process in Eq.(\ref{eq:coefficient}), we have to understand how to treat the sum over polarization of an off-shell gluon. Catani, Ciafaloni and Hautmann in \cite{Catani:1990eg} and \cite{Catani:1994sq} address this problem by defining
\begin{equation}
        d_{CH}^{\mu\nu}(k_1) = (d-2) \frac{ k_t^\mu k_t^\nu}{\textbf{k}_t^2}.
\end{equation}
This projector by construction selects the dominant part of the amplitude in the high-energy region and if we take its on-shell limit we get
\begin{equation}
        \lim_{\bar{k}_{\perp}^2\rightarrow 0 }\langle d_{CH}^{\mu \nu} \rangle = g_{\perp}^{\mu \nu},
\end{equation}
that is the usual sum over polarization of an on-shell gluon. This projector works well if we are interested in resumming LL terms, but we are investigating if it works also when we want to resum the following logarithmic order. An alternative definition for this tensor can be
\begin{equation}
    d^{\mu \nu}(k_1,n) = -g^{\mu \nu}+\frac{k_1^\mu n^\nu+k_1^\nu n^\mu}{k_1\cdot n}.
    \label{eq:dmunu}
\end{equation}
This tensor is derived by considering the off-shell incoming gluon as emitted by an on-shell massless quark. This is only one of the possible choices that one can make. For example, if the off-shell gluon comes from the splitting of an other gluon, we get a different result. We refer to our future work for a complete treatment of this problem. For the purpose of this work we use the definition in Eq.(\ref{eq:dmunu}) to give a more compact result. 

\subsection{Principal value prescription}

Since we worked in light-cone gauge, we found in our calculations some non-covariant loop integrals. These contributions arise because of the form of the gluon propagator in this gauge:
    \begin{equation}
        \Pi^{\mu \nu}_{a,b}(k,n) = \frac{\mathrm{i}\delta_{a,b}}{k^2}\left[-g^{\mu \nu}+\frac{k^\mu n^\nu+k^\nu n^\mu }{k\cdot n}\right].
    \end{equation}
Moreover, the second term in the brackets gives rise to some spurious singularities only due to the gauge choice. These singularities must be regulated and, since they are gauge-dependent, must not be present in the final result. 

We regulate these singularities with the Principal Value (PV) prescription    \cite{Konetschny:1975he,Pritchard:1978ts,Curci:1980uw}, while another possible prescription is the Mandelstam-Leibbrant (ML) \cite{mandelstam1983light,leibbrandt1984light}. Both these two possibilities have strengths and weaknesses, for a complete discussion on this topic see \cite{bassetto1991yang}.

The PV prescription consists in substituting
\begin{equation}
        \frac{1}{k\cdot n} \rightarrow \frac{k\cdot n}{(k\cdot n)^2+\delta^2(p\cdot n)},
\end{equation}
where the regulator is $\delta$ and $p^\mu$ is some external momenta (this term is there for dimensional reasons). Computed all the integrals with this regularization, we will take the limit where $\delta$ goes to zero. The final result for the complete coefficient function, where we put together both the real and the virtual contribution, must be independent on $\delta$ since it is gauge-invariant. This will provide a check for our calculation.

The most general loop integral that arises in our calculation is of the form 
\begin{equation}
    I_n = \int \frac{d^d k}{(2 \pi)^d}\frac{f\left(k\cdot n\right)}{D_1\, D_2\, \ldots\, D_n},
\end{equation}
where $D_i$ are the usual covariant denominators while all the non-covariant part is encoded in the function $f(k\cdot n)$. We derived some general results that allow us to compute non-covariant loop integrals without making assumptions on the non-covariant part. For a complete treatment of this topic we refer to our future work. The case where the numerator is a more complicated structure with different powers of the loop momentum can always be reduced to the scalar case. 

\subsection{Renormalization}
Another issue we encountered working in light-cone gauge is the definition of the ultraviolet counterterms. In fact, in this gauge, their structure is more complicated and cannot be trivially related to the structure of the terms in the Lagrangian. A discussion on this topic can be found in \cite{Curci:1980uw} or in \cite{bassetto1991yang}. 

Here we report, as an example, the result for the counterterm of the one-loop gluon propagator:
\begin{equation}
    \begin{aligned}
        &\Pi^{\mu\nu}(k_1,n) =- \mathrm{i} \frac{\alpha_s}{4 \pi}  \frac{C_A \delta_{a,b}}{\epsilon}\left[\left(\frac{11}{3} + 4\ln(\delta)\right)\left(k_1^\mu k_1^\nu- k_1^2 g^{\mu \nu}\right)\right. \\
        &\left. - 4\left(1+ \ln(\delta)\right)\left(k_1^\mu k_1^\nu- \frac{k_1^2}{k_1\cdot n} (k_1^\mu n^\nu + k_1^\nu n^\mu) + \frac{k_1^4}{(k_1\cdot n)^2}n^\mu n^\nu \right) \right].
    \end{aligned}
    \label{eq:CTgluonprop}
\end{equation}
The first thing we notice is the dependence on $\ln{\delta}$. This is due to the presence of spurious singularities and must vanish in the final result. Another observation one can make is about the tensorial structure. On the first row of Eq.(\ref{eq:CTgluonprop}) we recognize the tensorial structure of the counterterm for the gluon propagator that one can derive from the Lagrangian, while the second row is something new. It is however interesting to notice that the counterterm is still transverse with respect to the gluon momentum $k_1^\mu$.

\begin{figure}[h]
    \centering
    \begin{minipage}[c]{0.3 \textwidth}
     \centering
    \includegraphics[width=0.6 \linewidth]{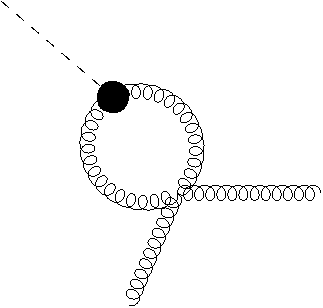}
    \end{minipage}
    \hspace{0.01 \textwidth}
    \begin{minipage}[c]{0.3 \textwidth}
     \centering
    \includegraphics[width=0.6 \linewidth]{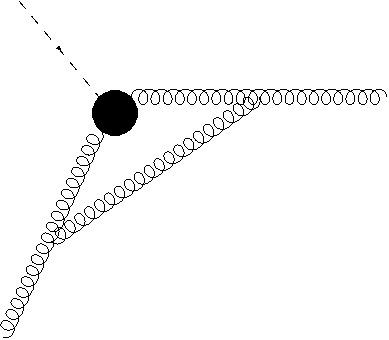}
    \end{minipage}
    \hspace{0.01 \textwidth}
    \begin{minipage}[c]{0.3 \textwidth}
     \centering
    \includegraphics[width=0.6 \linewidth]{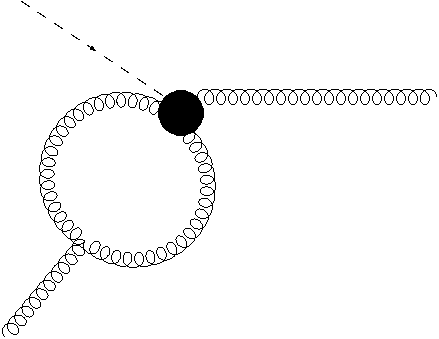}
    \end{minipage}
    \caption{Diagrams contributing to the one-loop correction to the gluon-Higgs vertex.}
    \label{fig:diagramsoneloop}
    \end{figure}

We also have to define the counterterm for the Higgs-gluon effective vertex. We choose to define them by computing the ultraviolet pole of each diagram that contributes to the virtual part of this coefficient function. To select only the UV pole, we compute the diagrams in Fig.\ref{fig:diagramsoneloop} with both the incoming and the outgoing gluons off-shell. In this case, the result is quite more complicated than the one in Eq.(\ref{eq:CTgluonprop}).

\subsection{Virtual contribution}
Here we present our first result. The virtual one-loop contribution to the coefficient function of the Higgs-induced DIS with the incoming gluon off-shell is
\begin{equation}
\begin{aligned}
    &\mathcal{C}^{(1)}_{g^* H\rightarrow g} =\frac{\ln \left(\xi \right) -\ln \left({1+\xi}\right)}{\epsilon} \frac{\left(\xi ^2+2 \xi +2\right) \left(\xi ^3-\xi ^2+\xi -1\right)  }{4 (\xi +1)^2}\\
    &+\frac{\xi  \left(2 \xi ^3-2 \xi -1\right) }{(\xi +1)^3}\ln^2(\delta) + \frac{\xi ^3  \ln (1+\xi )}{2 (\xi +1)^2}\ln(\delta)+\frac{1}{36 (\xi +1)^3} \left[33 \pi ^2 \xi ^4 \right.\\
    &\left.+33 \pi ^2 \xi ^3+62 \xi ^2-162 \xi ^4 \ln ^2(\xi )-9 \xi ^4 \ln (\xi )-117 \xi ^4 \ln ^2(\xi +1)-9 \xi ^2 \ln (\xi )\right.\\
    &\left.-162 \xi ^3 \ln ^2(\xi )-135 \xi ^3 \ln ^2(\xi +1)+27 \xi ^2 \ln ^2(\xi ) -36 \xi ^3 \ln (\xi ) -39 \pi ^2\right. \\
    &\left.-9 \xi ^2 \ln ^2(\xi +1)+162 \xi ^4 \ln (\xi -1) \ln (\xi )-54 \xi ^4 \ln (\xi -1)\ln (\xi +1)-39 \pi ^2 \xi \right.\\
    &\left. +180 \xi ^4 \ln (\xi ) \ln (\xi +1)+9 \xi ^4 \ln (\xi +1)+180 \xi ^3 \ln (\xi -1) \ln (\xi )-67 \xi^4\right.\\
    &\left.-72 \xi ^3 \ln (\xi -1) \ln (\xi +1)+198 \xi ^3 \ln (\xi ) \ln (\xi +1)+18 \xi ^2 \ln (\xi -1) \ln (\xi ) \right. \\
    &\left.-18 \xi ^2 \ln (\xi -1) \ln (\xi +1)-18 \xi ^2 \ln (\xi ) \ln (\xi +1)+18 \xi ^2 \ln (\xi +1)-67 \right.\\
    &\left.+342 \xi  \ln ^2(\xi )+234 \xi  \ln ^2(\xi +1)+315 \ln ^2(\xi )+225 \ln ^2(\xi +1)+18 \xi  \ln (\xi )\right.\\
    &\left.-180 \xi  \ln (\xi -1) \ln (\xi )+72 \xi  \ln (\xi -1) \ln (\xi +1)-360 \xi  \ln (\xi ) \ln (\xi +1)\right.\\
    &\left.-18 \xi  \ln (\xi +1)-180 \ln (\xi -1) \ln (\xi )+72 \ln (\xi -1) \ln (\xi +1)-9 \ln (\xi +1)\right.\\
    &\left.-324 \ln (\xi ) \ln (\xi +1)+36 \xi ^3 \ln (\xi +1)-36 \left(\xi ^4+\xi ^3-\xi ^2+1\right) \text{Li}_2\left(\frac{1}{\xi }\right) \right.\\
    &\left.-18 \left(2 \xi ^4+2 \xi ^3+\xi ^2-3 \xi -4\right) \text{Li}_2\left(\frac{1}{\xi ^2}\right)\right],
\end{aligned}
\label{eq:coefficient}
\end{equation}
where $\xi = \frac{k_t^2}{Q^2}$ and we dropped the overall factor $C=-\frac{2\alpha_s^3 \sqrt{2} G_F Q^2}{3\,\pi^3}$, where $Q^2$ is the virtuality of the Higgs Boson. We notice that in this result, after the renormalization, there is a residual infrared singularity. This singularity must vanish once we put together the virtual and the real contribution and it will be an important cross-check. We also notice that, while in the coefficient of the pole there is no dependence on the regularization parameter $\delta$, in the finite part we still have some $\ln(\delta)$. These logs must cancel once we put together the virtual and the real contribution and will provide another cross-check. 

\section{Outlooks}

We computed the virtual one-loop contribution to the Higgs-induced DIS with the incoming gluon off-shell in light-cone gauge. We particularly focused on the understanding of light-cone gauge and its properties. We introduced a way to deal with spurious singularities and non-covariant integrals. 

To give a complete result, that will be useful in the resummation of the coefficient function of this process up to NLL accuracy, there are still some steps that have to be done that will be presented in following works. First of all, we have to compute the real contribution to this process which will also allow us to do many cross-checks. Moreover, we have to understand how to deal with the sum over polarization of the incoming gluon that is off-shell. We aim to write it in a process-independent way that will select the dominant part in the high-energy region and will give the correct result in the on-shell limit. 


\bibliography{Higgs_DIS}
\bibliographystyle{ieeetr}

\end{document}